\documentclass[12pt]{article}
\usepackage{epsfig}
\begin{document}
\title{\bf{Precise Variational Calculation For The Doubly Excited State $(2p^2)^3P^e$ of Helium.}}
\author{Tapan K. Mukherjee\footnote {Current Address: Department of Physics,
Sikkim Manipal Institute of Technology,
Majitar, Rangpo, East Sikkim-737132, India.} and Prasanta K. Mukherjee\\
Department of Spectroscopy\\
Indian Association for the Cultivation of Science\\
Jadavpur, Kolkata 700 032, India}
\date{}
\maketitle

\begin{abstract}
Highly precise variational calculations of non-relativistic
energies of the $(2p^2)^3P^e$ state of Helium atom are
presented.We get an upper bound energy $ E=-0.71050015565678 $
a.u.,the lowest yet obtained.
\end{abstract}

\section{Introduction}
Multiply excited energy states in atoms have been the subject of
many experimental and theoretical investigations. Specially doubly
excited states (DES) for neutral helium is particularly important
as the DES provide a fundamental testing ground of the accuracy of
the theoretical treatment. The review article of $Hol{\o}ien$$^1$
and $Fano^2$ gives a detailed list of references.\\
A large number of DES of neutral helium can auto-ionize to the
continuum above the $1s$$(^2S)$ ground state of $He^+$. Besides,
there are also non-auto-ionizing doubly excited states in helium.
These exactly quantized DES states decay to the lower excited
states through electric dipole interaction giving rise to sharp
spectral line. It is worthwhile to mention that relativistically
these states may auto-ionize, but the auto-ionization life times
are still appreciably longer than the mean radiative life times of
the allowed electric
dipole $transition^1$. \\
$Compton$ and $Boyce^3$, $Kruger^4$, $Whiddington$ and
$Priestley^5$ are the pioneers to observe such exactly quantized
DES of helium. Among such non-auto-ionizing doubly excited levels
in helium, the (2$p^2$)$^3P^e$ state is the lowest lying P-state
of even parity. The interpretation of $320.392A^0$ line in helium
as the $(2p^2)^3P^e\rightarrow(1s2p)^3P^0$-transition by
$Kruger^4$ was later confirmed by $Wu^6$ on the basis of
theoretical calculation.
 $Tech$ and $Ward^7$
reinvestigated the matter and performed highly accurate
spectroscopic measurement of the line
$2p^2(^3P^e)\rightarrow(1s2p)^3P^0$ at $320.2926\pm0.0010A^0$. A
discrepancy of about $100cm^{-1}$ between the measurements of
$Tech. et.al^7$ and that of $Kruger^4$ is due to the
unavailability of accurate standard wave-lengths at the time of
measurement of $Kruger^4$. $Berry$ $et. al^8$ also observed the
line $2p^2(^3P^e)\rightarrow1s2p(^3P^0)$ at
$320.40\pm0.3A^0$.\\
$Drake$ and $Dalgarno^9$, $Hol{\o}ien^{10}$, $Bhatia^{11}$
calculated the energy of the $(2p^2)^3P^e$- state of helium by
using variational method. Using variational perturbation method
$Aashamar^{12}$ obtained the accurate eigenvalue for the
$(2p^2)^3P^e$- state of neutral helium. All these theoretical
investigations predicted the wave length of
$2p^2(^3P^e)\rightarrow1s2p(^3P^0)$ transition of helium range
from $320.288$ to $320.293A^0$, and hence are in
agreement with the measurement of $Tech$ $et. al^7$\\
Under such circumstances it is necessary to perform very accurate
spectroscopic calculation of the $(2p^2)^3P^e$-state of helium. In
this article, we propose a method to improve the DES wave function
to get not only the best upper bound energy but also rapid
convergence with respect to the size of the trial space.
\section{Theory}
The rotational invariance of the Hamiltonian makes it possible to
express the variational equation of two electrons in the field of
a fixed nucleus in terms of three independent $variables^{13}$.The
three coordinates are the sides of the triangle formed by the
three particles i.e., two electrons and the fixed nucleus. The
reduction of the three $Eulerian$ angles, defining the orientation
of this triangle in space, from the variational equation is an
immediate consequence of the spherical symmetry of the field. For
any $^3P$- state of even parity arising from two equivalent
electrons, the general variational equation (10) of $Ref.13$
reduces to
\begin{eqnarray}
\delta\int[\left({\partial f ^0_1\over \partial r_1}\right)^2 +
\left({\partial f ^0_1\over \partial r_2}\right)^2 +
(r^{-2}_1+r^{-2}_2)\left({\partial f ^0_1\over \partial
\theta_{12}}\right)^2 +(r^{-2}_1+r^{-2}_2)(f^0_1)^2
{1\over\sin^2\theta_{12}}\nonumber\\
+2(V-E)(f^0_1)^2]dV_{r_1,r_2;\theta_{12}}=0
\end{eqnarray}
subject to the normalization condition
\begin{equation}
\int(f^0_1)^2 dV_{r_1,r_2;\theta_{12}}=1
\end{equation}
The symbols in equation (1) and (2) are same as of $Ref.13$. We
use
atomic units throughout.\\
The correlated wavefunction is given by,
\begin{eqnarray}
f^0_1(r_1,r_2,r_{12})=\eta_1(1)\eta_1(2)[\sum_{l>0}\sum_{m>0}
\sum_{n\geq0}B_{lmn}r_1^lr_2^mr_{12}^n\sin\theta_{12} + Exchange]\nonumber\\
+[\eta_1(1)\eta_2(2)\sum_{l>0}\sum_{m>0}\sum_{n\geq0}C_{lmn}r_1^lr_2^mr_{12}^n\sin\theta_{12}+
Exchange]\nonumber\\
+\eta_2(1)\eta_2(2)[\sum_{l>0}\sum_{m>0}\sum_{n\geq0}D_{lmn}r_1^lr_2^mr_{12}^n\sin\theta_{12}+
Exchange]
\end{eqnarray}
where, $\eta\sim e^{-\rho r}$
 and $\rho$ is the non-linear parameter. The linear coefficients
 B, C, D along with energy eigenvalue E are determined by matrix
 diagonalisation method.
\section{Results}
The results of our calculation are given in $Table1$. All
calculations were carried out in quadruple precision. The orbital
exponents $\rho_1$ and $\rho_2$ were optimized using the
$Nelder-Mead^{14}$ procedure and are given in $Table1$.\\
There are other variational $ calculations^{9-11} $ of the
$(2p^2)^3P^e$-states for $Helium$. So far best result was obtained
by $Bhatia^{11}$. A comparison of present results to that of
$Bhatia^{11}$ is given in $Table1$. The total number of terms (N)
for each calculation is given in the first column. The results
obtained by $Bhatia^{11}$ in second column are compared with the
present results in the last column. It's remarkable that for a
given number of term (N) the results of the present basis sets are
better than that of $Bhatia^{11}$ for next largest basis set e.g.
our result for $21$ parameter calculation is better than that of
35 parameter calculation of $Bhatia^{11}$, again our 84 parameter
result is better than the $97$ parameter result of $Bhatia^{11}$
as is evident from $Table 1$. Substantial reduction of the number
of terms i.e. the basis set size is a clear advantage of the
present method for a given calculated energy value.
\begin{table}
\begin{center}
\caption {\rm {\bf{Non-relativistic energy (-E) of $He$ atom in
$(2p^2)^3P^e$-state. All energies are in a. u. The non linear
parameters are $\rho_1=0.81006481$ and $\rho_2=1.07917071$.}}}
\begin{tabular}{l l l }\\
\hline\hline
 N      &$Bhatia$    & $Present Method$\\
\hline\\

20  & 0.710456705905 &                 \\
21  &                & 0.71049962427254\\
35  & 0.710497876335 &                 \\
39  &                & 0.71050006887316\\
54  &                & 0.71050014164022\\
56  & 0.710500049935 &                 \\
66  &                & 0.71050014872252\\
70  & 0.710500140510 &\\
84  & 0.710500142765 & 0.71050015218194\\
90  & 0.710500149950 &\\
95  & 0.710500151000 &\\
96  & 0.710500151515 &\\
97  & 0.710500152070 &\\
99  &                & 0.71050015410607\\
150 &                & 0.71050015526295\\
210 &                & 0.71050015554599\\
267 &                & 0.71050015561803\\
300 &                & 0.71050015564129\\
321 &                & 0.71050015564988\\
336 &                & 0.71050015565678\\\\
\hline
\end{tabular}
\end{center}
\end{table}
It is relevant to mention that using variational-perturbation
technique $Aashamar^{12}$ obtained the non-relativistic energy for
$(2p^2)^3P^e$- state of helium as $0.71050015560$$a.u.$. Present
variational results of $0.710500155678$$a.u.$ is even slightly
lower than that of
variational-perturbation $results^{12}$.\\
$Tech$ $et.$ $al^7$ observed the sharp line in far ultra-violet
region with wave number $312,214.52\pm0.97$ $cm^{-1}$
corresponding to a wavelength $320.2926\pm0.0010$ $A^0$. In order
to compare with experiment, we use the well known
$experimental^{15}$ line with wave number
$169,087.01\pm0.15$$cm^{-1}$ corresponding to the transition
$(1s2p)^3P^0\rightarrow1s^2(^1S)$. Adding the wave number of the
above two lines, an experimental value of
$481,301.53\pm0.98$$cm^{-1}$ for the energy of the doubly excited
$(2p^2)^3P^e$ term relative to the ground $(1s^2)^1S$ results.
Similarly, combining the observation of $Kruger^4$ and the
experiment of $Martin^{15}$, an experimental value of
$481,205cm^{-1}$ for the energy of the $(2p^2)^3P^e$-term relative
to the ground $1s^2(^1S)$ results. It is clear that there is a
discrepancy of approximately 100$cm^{-1}$ between the experimental
results of $Kruger^4$ and $Tech^7$. We obtained the position of
the $(2p^2)^3P^e$-state above the ground $1s^2(^1S_0)$-state by
subtracting our calculated energy for the $(2p^2)^3P^e$-state from
the well known$^7$ energy $-637,219.54cm^{-1}$ of the ground
$1s^2(^1S_0)$ state of helium. The conversion factor
$1a.u.=219,444.528cm^{-1}$ is used. Our calculated value of
$481,304.17cm^{-1}$ for the energy of $(2p^2)^3P^e$-state above
the ground $(1s^2)^1S_0$-state is far from the experimental value
of $481,205cm^{-1}$ of $Krugar^4$, but agrees fairly well with the
experimental value of $481,301.53\pm0.98cm^{-1}$ of $Tech$ $et.
al.^7$. $Aashamar$ obtained a value of $481,301.62cm^{-1}$ for the
same including mass polarization, relativistic and radiative
effects. A difference of approximately $3cm^{-1}$ between our
non-relativistic results and that of $Aashamar$ or experiment is
due to relativistic and other correction.\\
Finally taking the difference between the present calculated wave
number $481,304.17cm^{-1}$ for the $(2p^2)^3P^e$-state above the
ground $1s^2(^1S_0)$-state of helium and the experimental$^{15}$
line at wave number $169,087cm^{-1}$ for the transition
$1s2p(^3P^0)\rightarrow1s^2(^1S_0)$ of helium, we get the wave
number $312,217.16cm^{-1}$, corresponding to a wavelength
$320.2899^0A$ for the transition
$2p^2(^3P^e)\rightarrow1s2p(^3P^0)$ of helium.
\section{Acknowledgement}
We express gratitude to $Professor A. K. Bhatia$ for helpful
suggestions. The authors are thankful to $Department$ of $Atomic$
$Energy$ ($DAE$), $Govt.$ of $India$ for a research grant $No.
2003/37/12/BRNS/622$.
\section{References}
$^{1}$ E. Hol{\o}ien, Nucl.Instrum. Methods \underline{90},
229(1970)\\
$^{2}$ U. Fano, in Atomic Physics, edited by V. W. Huges, B.
Bedeson, V. W. Cohen and F. M. J. Pichanick(Pleunum, New York,
1969) p.209\\
$^{3}$ K. T. Compton and J. C. Boyce, J. Franklin Inst, \underline{205}, 497(1928).\\
$^{4}$ P. G. Kruger, Phys. Rev. \underline{36}, 855(1930).\\
$^{5}$ R. Whiddington and H. Priestley, Proc. Roy. Soc.(London)A
\underline{145}, 462(1934)\\
$^{6}$ T. Y. Wu, Phys. Rev. \underline{66}, 291(1944).\\
$^{7}$ J. L. Tech and J. F. Ward, Phys. Rev. Lett. \underline{27}, 367(1971).\\
$^{8}$ H. G. Berry, I. Martinson, L. J. Curtis and L. Lundin, Phys. Rev. A \underline{3}, 1934(1971)\\
$^{9}$ G. W. F. Drake and A. Dalgarno, Phys. Rev. A \underline1, 1325 (1970)\\
$^{10}$ E. Hol{\o}ien, J. Chem. Phys \underline{29}, 676(1958);
and Phys. Norvegica \underline{1}, 53(1960)\\
$^{11}$ A. K. Bhatia, Phys. Rev.A. \underline{2}, 1667(1970).\\
$^{12}$ K. Aashamar, Institute for Theoretical Phys., University
of Oslo, Norway,
Institute Report No.35,1969\\
$^{13}$ Tapan K. Mukherjee and Prasanta K. Mukherjee, Phys. Rev.A
\underline{50}, 850(1994).\\
$^{14}$ J. A. Nelder and R. Mead, Computer J. \underline {7} , 308
(1965).\\
$^{15}$ W. C. Martin, J. Res. Nat. Bur. Stand. Sect. A
\underline{64}, 19(1960)\\
\end{document}